\documentclass{WileyMSP-template}
\usepackage{amsmath}
\usepackage{subcaption} 
\usepackage{caption} 
\usepackage{float}
\usepackage{array,multirow}
\usepackage{booktabs}
\usepackage{nameref}
\usepackage{xcolor,colortbl}
\usepackage{url}
\usepackage[T1]{fontenc}

\begin{document}


\title{3D ferroelectric phase field simulations of polycrystalline multi-phase hafnia and zirconia based ultra-thin films}

\maketitle


\author{Prabhat Kumar*}
\author{Michael Hoffmann}
\author{Andy Nonaka}
\author{Sayeef Salahuddin}
\author{Zhi (Jackie) Yao*}



\begin{affiliations}
Prabhat Kumar, Andrew Nonaka, Zhi (Jackie) Yao\\
Applied Mathematics and Computational Research Division,\\ Lawrence Berkeley National Laboratory,\\ CA, 94720, USA\\

E-mail: prabhatkumar@lbl.gov\\
E-mail: jackie\_jhiyao@lbl.gov \\

Michael Hoffmann, Sayeef Salahuddin\\
Department of Electrical Engineering and Computer Sciences, \\ University of California, \\ Berkeley, CA, 94720, USA

\end{affiliations}



\begin{abstract}

HfO$_2$- and ZrO$_2$-based ferroelectric thin films have emerged as promising candidates for the gate oxides of next generation electronic devices. Recent work has experimentally demonstrated that a tetragonal/orthorhombic (t/o-) phase mixture with partially in-plane polarization can lead to negative capacitance (NC) stabilization. However, there is a discrepancy between experiments and the theoretical understanding of domain formation and domain wall motion in these multi-phase, polycrystalline materials. Furthermore, the effect of anisotropic domain wall coupling on NC has not been studied so far. Here we apply 3D phase field simulations of HfO$_2$- and ZrO$_2$-based mixed-phase ultra-thin  films on silicon to understand the necessary and beneficial conditions for NC stabilization. We find that smaller ferroelectric grains and a larger angle of the polar axis with respect to the out-of-plane direction enhances the NC effect. Furthermore, we show that theoretically predicted negative domain wall coupling even along only one axis prevents NC stabilization. Therefore, we conclude that topological domain walls play a critical role in experimentally observed NC phenomena in HfO$_2$- and ZrO$_2$-based ferroelectrics.

\end{abstract}


\section{Introduction}

Recently, ferroelectric thin films have emerged as promising materials for high performance and energy-efficient nanoelectronic devices with applications in computation and data storage \cite{Khan2020,Mikolajick2021NextGF}. By applying an electric field to the ferroelectric film, its spontaneous polarization can be switched thus enabling low-power non-volatile memory devices suitable for digital storage. Furthermore, ferroelectric materials are promising for neuromorphic computing due to their ability to provide analog storage options that can be dynamically changed for learning and adaptation \cite{Oh2019,Miko2023}. In ultra-thin ferroelectric films, depolarization fields lead to a suppression of the overall polarization, typically through the formation of ferroelectric domains \cite{Mehta1973DepolarizationFI,Wurfel1973DepolarizationFieldInducedII,Bratkovsky1999AbruptAO}. Interestingly, such depolarized ferroelectrics can exhibit an effective negative permittivity and thus a negative capacitance (NC) \cite{Landauer1976,Bratkovsky2006DepolarizingFA}, which is of interest for overcoming the fundamental limitations of performance and energy-efficiency of nanoscale transistors \cite{salahuddin2008use}. Thus, the integration of ferroelectric gate oxides into transistors enables both more energy-efficient memories as well as logic devices with lower operating voltage, higher drive current, and reduced short channel effects \cite{Khan2020,Hoffmann2021FerroelectricGO}.
\vspace{0.6em}

For practical device applications, fluorite-structure ferroelectrics based on HfO$_2$ and ZrO$_2$ 
are most promising due to their compatibility with standard semiconductor manufacturing processes as well as their scalability to ultra-thin thickness \cite{Boscke2011,Muller2012simple,Mikolajick2021NextGF,Silva2023RoadmapOF}. When integrated into a device such as a transistor, these fluorite-structure ferroelectrics exhibit a polycrystalline and multi-phase film morphology, as indicated in Fig. 1(a). Due to their similar formation energies, both the polar orthorhombic Pca2$_1$ (o-)phase and the non-polar tetragonal P4$_2$/nmc (t-)phase are often found together in polycrystalline HfO$_2$ and ZrO$_2$ based films \cite{Materlik2015}. In addition, amorphous and non-polar monoclinic P2$_1$/c phase fractions might occur when the film thickness is decreased or increased, respectively \cite{Hoffmann2015stabilizing}. In the present study, we do not consider the monoclinic phase since it is typically absent in ultra-thin films, where surface energy effects reduce its stability compared to the t- and o-phase \cite{Materlik2015}. Since the angle between the polar axis and the film plane varies for each ferroelectric grain, the multi-phase polycrystalline nature of these films can have a strong effect on domain formation and the macroscopic device characteristics \cite{Schroeder2022TheFA,Pujar2023PhasesIH}. 
Recently, it has been demonstrated that ultra-thin HfO$_2$ and ZrO$_2$ based films with a thickness of as low as 0.5 nm to 2 nm can retain their ferroelectric or antiferroelectric behavior, which is of interest for ultimately scaled electronic devices \cite{Cheema2020EnhancedFI,Lee2021UnveilingTO,cheema2022emergent}. Indeed, it was found that ferroelectricity is enhanced when the film thickness is reduced to 2 nm and below, which is opposite to the size effect observed in classical ferroelectrics such as perovskites \cite{Cheema2020EnhancedFI}. Furthermore, when grown on single-crystalline silicon with a thin SiO$_2$ layer, these ultra-thin HfO$_2$ and ZrO$_2$ based ferroelectrics exhibit a preferential texture, where the polar axis is partially in the film plane \cite{Cheema2020EnhancedFI,Lee2021UnveilingTO}. Interestingly, it was found that 1-2 nm thick HfO$_2$ and ZrO$_2$ based films which consist of a o/t-phase mixture when grown on silicon, show an effective negative capacitance, thus enabling a reduction of the equivalent oxide thickness (EOT) below the physical thickness of the SiO$_2$ interfacial layer \cite{cheema2022ultrathin,Jo2023NegativeDC}. The EOT of a gate stack is defined as the thickness of a SiO$_2$ layer with relative permittivity of 3.9, which has the same capacitance as that gate stack. NC transistors with reduced EOT exhibit improved device performance and can operate at lower gate voltages, as indicated in Fig. 1(b). A further reduction of the EOT without compromising the gate oxide reliability and leakage current is one of the major challenges in making transistors smaller \cite{hoffmann2021}. However, the microscopic origin of this NC effect in mixed phase ferroelectric HfO$_2$ and ZrO$_2$ based ultra-thin films is not fully understood yet \cite{Hoffmann2020whats}.
\vspace{0.6em}

While an \textit{in situ} experimental characterization of the domain patterns in such ultra-thin mixed phase ferroelectric films is difficult, numerical simulations can give helpful insight into the possible evolution of domain patterns and their relationship to NC \cite{Park2019ModelingON,Saha2019MultiDomainNC}. Density functional theory (DFT) has been applied to study the stability and formation energies of various domain wall configurations of HfO$_2$ and ZrO$_2$ based ferroelectrics \cite{Ding2020TheAD,Lee2020ScalefreeFI}. It was found that abrupt 180$^\circ$ domain walls can form easily due to a slightly negative domain wall energy along one of the axes perpendicular to the polar axis of the orthorhombic crystal \cite{Lee2020ScalefreeFI}. In addition, it was found using DFT that there is a further topological class of 180$^\circ$ domain walls, which could explain the experimentally observed low switching barriers as well as improved domain wall mobility in films with t/o-phase mixture \cite{Choe2021UnexpectedlyLB}.
While DFT calculations are useful for studying the fundamental properties of domain walls in the ferroelectric o-phase in HfO$_2$ and ZrO$_2$ based materials, it cannot give insight into the mesoscopic and dynamic behavior of domains and domain walls in polycrystalline mixed-phase films. For these more practically relevant film properties and length scales, ferroelectric phase field simulations have proved useful \cite{Koduru2021VariationAS,Park2022NegativeCF}. NC phenomena in ferroelectrics have been studied using phase field simulations mainly in 2D \cite{Park2019ModelingON,Saha2019MultiDomainNC,Park2022NegativeCF}. However, the strongly anisotropic domain wall coupling in HfO$_2$ and ZrO$_2$ based ferroelectrics predicted by DFT requires a 3D simulation to fully capture the domain structures and dynamics \cite{Paul2022DirectionDependentLD}. 
\vspace{0.6em}

Here, we investigate the 3D domain pattern formation and associated NC effects in polycrystalline, multi-phase ultra-thin HfO$_2$ and ZrO$_2$ based ferroelectrics on silicon, by applying FerroX \cite{Kumar2023}, a 3D phase field simulation tool. To study these effects, metal-ferroelectric-insulator-semiconductor-metal (MFISM) capacitor structures similar to Fig. 1(c) are simulated in 3D, where a ferroelectric o-phase grain is surrounded in the film plane by the non-ferroelectric t-phase. The capacitance per ferroelectric grain area is then compared to the capacitance of the structure without the mixed phase ferroelectric film (metal-insulator-semiconductor-metal (MISM) structure), see Fig. 1(d). The capacitance enhancement of the MFISM structure compared to the MISM case is used to quantify the ferroelectric NC effect. We further vary the ferroelectric grain size, orientation, and the anisotropic domain wall coupling constants to better understand their relationship to the NC effect. These insights will help gain more insight into the origin of NC in such ultra-thin ferroelectrics and how to tailor their properties for further improvements in device behavior. 

\begin{figure}[h]
    \centering
    \includegraphics[width=12cm]{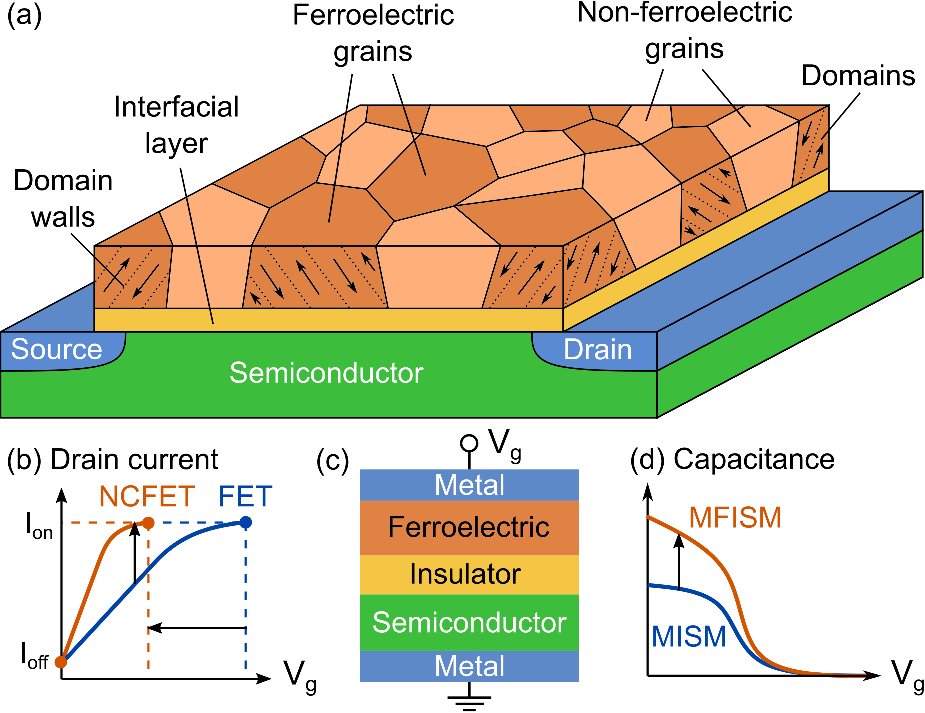}
    \caption{(a) Schematic structure of a field-effect transistor (FET) with ultra-thin HfO$_2$/ZrO$_2$-based ferroelectric gate oxide. The polycrystalline multi-phase and multi-domain ferroelectric film  exhibits an effective negative capacitance (NC). (b) Logarithm of the drain current as a function of the gate voltage $V_g$ for a conventional FET and a negative capacitance FET (NCFET). For a matched Off-current $I_{off}$ the supply voltage can be reduced or the On-current $I_{on}$ can be boosted at the same supply voltage. (c) Simplified metal-ferroelectric-insulator-semiconductor-metal (MFISM) capacitor structure simulated in this study. (d) Due to the NC effect, the MFISM capacitor shows an overall capacitance enhancement compared to an otherwise identical metal-insulator-semiconductor-metal (MISM) structure.}
    \label{fig:1}
\end{figure}

\newpage

\section{Results and Discussion}
\subsection{Domain Dynamics in MFISM Structure}
To study the origins of the negative capacitance effect and its implications in the operation of transistors with ferroelectric gate oxides, we consider the MFISM heterostructure and compare it with an otherwise identical MSIM capacitor. The baseline MFISM structure consists of a 5.0 nm thick p-doped Silicon as the semiconductor layer, a 0.8 nm SiO$_2$ as the interfacial layer, and a 2.2 nm thick polycrystalline ferroelectric layer. The lateral dimensions of the films are 16.128 nm. These ferroelectric and dielectric layer thicknesses are similar to recent experimental reports of NC \cite{cheema2022ultrathin,Jo2023NegativeDC}. The ferroelectric layer consists of a t/o-phase mixture with partially in-plane polarization. To capture the texture of ultra-thin HfO$_2$ and ZrO$_2$ based ferroelectrics in our 3D phase field model, we introduced polar angles $(\theta_x, \theta_y , \theta_z)$, where $\theta_x = \theta_y = 0 \degree$ and $\theta_y = 45\degree$ correspond to the experimentally observed preferential $(111)$ texture \cite{cheema2022ultrathin}. The o-phase grain in the ferroelectric layer is for now assumed to be 10 nm $\times$ 10 nm in the center of the film and is surrounded on all sides by the non-polar t-phase in the rest of the film. Additional details of the modeling procedure are described in the Computational Section. Landau free energy coefficients and the domain wall coupling parameters are listed in Table~\ref{tab:sim_param} 
along with other numerical and physical parameters. We also performed a series of simulations of the MISM heterostructure (i.e. without the ferroelectric layer) with different insulator thicknesses while keeping all other parameters the same as in the MFISM case. These simulations enable us to quantify the NC behavior of the ferroelectric layer by comparing the capacitances of the MFISM and MISM stacks, denoted as $C_{\rm{MFISM}}$ and $C_{\rm{MISM}}$ respectively. The capacitance of the MFISM stack represents the combined capacitance of the ferroelectric, SiO$_2$ and silicon layers in series. Therefore, if $C_{\rm{MFISM}} > C_{\rm{MISM}}$, it suggests the manifestation of NC attributable to the presence of the ferroelectric layer. Furthermore, by comparing the voltage-dependent $C_{\rm{MFISM}}$ to $C_{\rm{MISM}}$ curves with different SiO$_2$ layers, we can directly determine the EOT of the MFISM stack as the SiO$_2$ thickness of the matching $C_{\rm{MISM}}$ curve \cite{Hoffmann2022iedm}.

\begin{figure}[h]
    \centering
    \includegraphics[width=\linewidth]{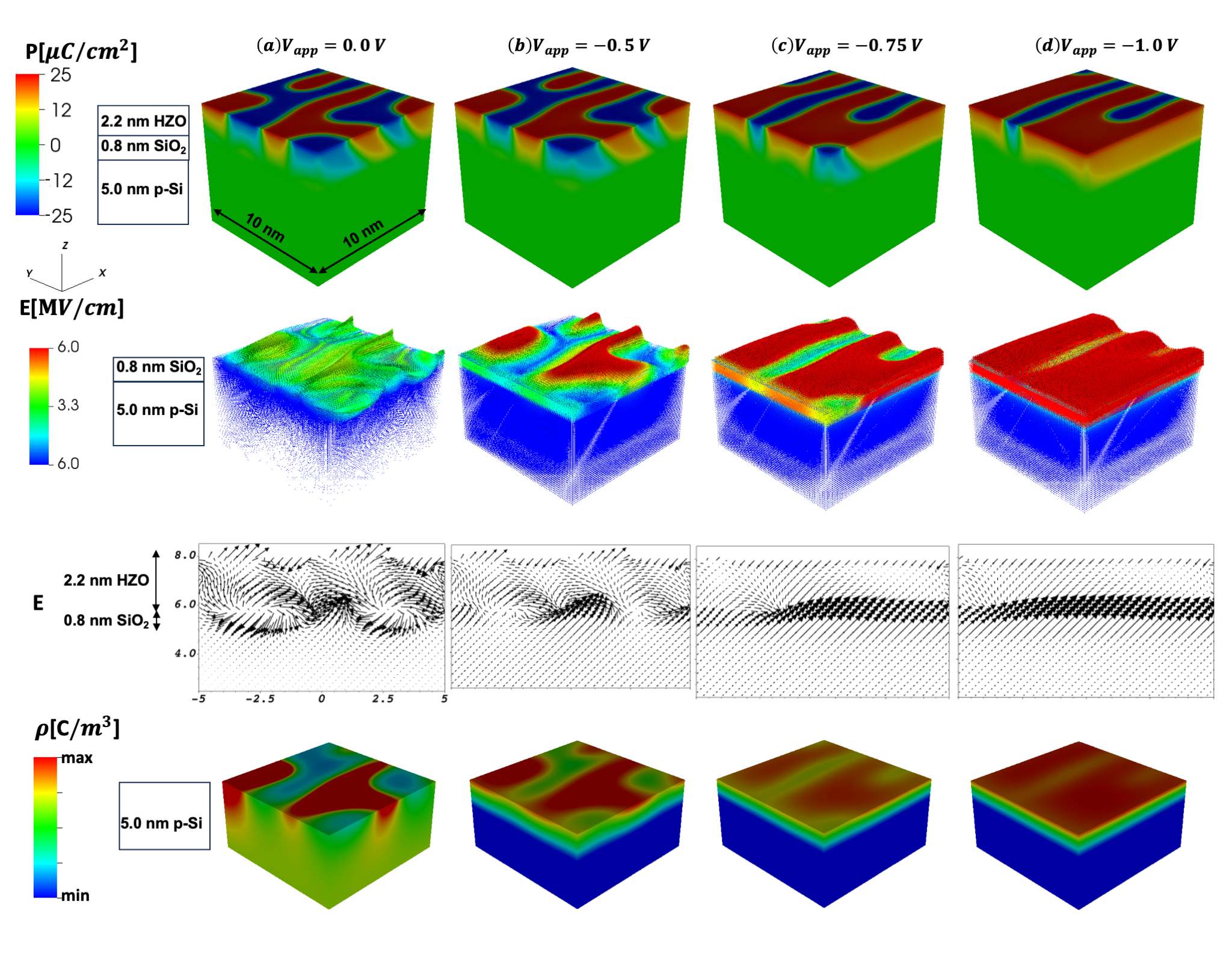}
    \caption{3D distribution of local Polarization, electric fields, and charge density for different applied voltages. Simulations use parameters listed in Table 1. Only the orthorhombic grain is shown in the top row, since the polarization in the surrounding t-phase is zero. Second row shows the distribution of electric field. To clearly show the direction of the electric field, a slice in the x-z plane at 1 nm from the t-o phase boundary is shown in the third row. The fourth row shows the distribution of charge density in the semiconductor region.}
    \label{fig:2}
\end{figure}

\begin{table}[h]
\centering
\begin{tabular}{l  l  r}\toprule
Parameters & Values & Units \\
\\[-1.8em]
\hline
 Landau free-energy coefficients: & & \\
 \\[-2.0em]
 $\alpha$ & $-2.5\times 10^9$ & $~\rm{Vm/C}$ \\
\\[-2.0em]
    $\beta$ & $6.0\times 10^{10}$ & $~\rm{Vm^5/C^3}$\\
\\[-2.0em]
$\gamma$ & $1.5\times 10^{11}$ & $~\rm{Vm^9/C^5}$\\
Gradient-energy coefficients $(g_{x} = g_{y} = g_z)$ & $1.0\times 10^{-10}$ & $~\rm{Vm^3/C}$\\
\\[-2.0em]
Polar axis angles $(\theta_x, \theta_y, \theta_z)$ & $(0, 45, 0)$ & degrees \\ 
\\[-2.0em]
Length $(L_o) = $ Width $(W_o)$ of o-phase & $10.0$ & nm \\ 
\\[-2.0em]
Length $(L_{\rm{F}}) = $ Width $(W_{\rm{F}})$ of the film & $16.128$ & nm \\ 
\\[-2.0em]
Thickness of the ferroelectric layer $(t_{\rm{F}}) $ & $2.2$ & nm \\ 
\\[-2.0em]
Thickness of the dielectric layer $(t_{\rm{I}})$ & $0.8$ & nm \\ 
\\[-2.0em]
Thickness of Si layer $(t_{\rm{S}})$ & $5.0$ & nm \\ 
\\[-2.0em]
Permittivity of the ferroelectric layer $(\epsilon_\mathrm{\rm F})$ & 24.0~(\rm{Hf$_{0.5}$Zr$_{0.5}$O$_2$}) & 1\\
\\[-2.0em]    
 Permittivity of the dielectric layer $(\epsilon_\mathrm{I})$ & $3.9~(\rm{SiO}_2)$ & 1\\
\\[-2.0em]
Permittivity of the Si layer $(\epsilon_\mathrm{S})$ & $11.7~(\rm{Si})$ & 1\\
\\[-2.0em]
Kinetic Coefficient $(\Gamma)$ & $100.0$ & $1/(\Omega \cdot \mathrm{m})$ \\
\\[-2.0em]
Grid Sizes $(\Delta x = \Delta y = \Delta z)$ & $0.125$  & $~\rm{nm}$ \\
\\[-2.0em]
Polarization penetration length $(\lambda)$ & $0.3$ & $\mathrm{nm}$ \\
\\[-1.8em]
Time step size $(\Delta t)$ & $0.25\times 10^{-13}$ & $~\rm{s}$ \\
[-0.3em]\bottomrule
\end{tabular}
\caption{Physical and numerical parameters used in the simulations}
\label{tab:sim_param}
\end{table}

\begin{figure}[h]
    \centering
    \includegraphics[width=0.45\linewidth]{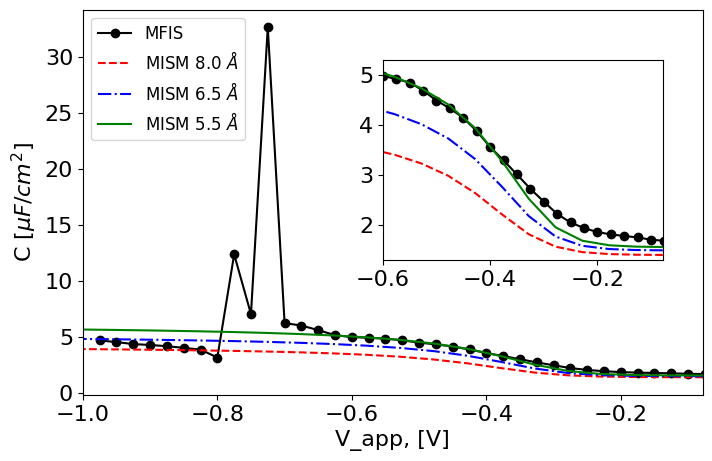}
    \caption{$C-V$ curves without ferroelectric (solid lines) and with ferroelectric + $8.0~\AA$ SiO$_2$ (black solid line with circles) showing the achievable EOT as low as $5.5~\rm{\AA}$. Simulations use parameters listed in Table 1. }
    \label{fig:3}
\end{figure}

Figure~\ref{fig:2} shows the 3D distributions of ferroelectric polarization $P$, electric field vectors $E$ 
, and free charge carrier density $\rho$ as a function of external applied voltage $V_{app}$ for the parameters in Table 1. Due to the ultra-thin ferroelectric film thickness and the unscreened bound polarization charge at the ferroelectric/SiO$_2$ interface, a 3D domain pattern forms to reduce the depolarization energy in the ferroelectric. The multi-domain patterns show a $45\degree$ tilt in the $x-z$ plane, due to the assumed (111) texture of the film. For an applied voltage $V_{app} = 0~V$, the negatively and positively polarized domains (blue and red) have roughly identical sizes, however, as $V_{app}$ is decreased (Fig. 2(b)), the positive domains become progressively larger by domain wall motion to align with the external electric field. The electric field vector plots indicate the presence of both in-plane and out-of-plane components, especially close to the domain walls. The tilted multi-domain structure results in inhomogeneous stray fields, which change significantly as the domain walls move and contribute to NC effects in these films \cite{Park2022NegativeCF}. The presence of t-phase surrounding the ferroelectric grain from all sides, leads to additional stray fields close to
the t/o-phase boundaries due to uncompensated in-plane polarization contribution. Note that although the domain walls have moved, the overall domain topology and number of domains for $V_{app} = -0.5~V$ remains the same compared to the initial $0~V$ case. This means, that increasing the voltage back from $-0.5~V$ to $0~V$ would lead to the same initial state, i.e. the domain wall motion in this voltage range is reversible.

However, there is a qualitative difference when comparing the domain patterns at $V_{app} = -0.5~V$ and $V_{app} = -0.75~V$: Some previously individual red domains coalesce along the x-axis into larger domains, when the applied voltage is reduced to $V_{app} = -0.75~V$. This is equivalent to a partially irreversible switching of the polarization of the blue domains, which previously separated the red domains. After this switching event at $V_{app} = -0.75~V$, the number of domains and overall domain topology has changed and cannot be recovered by simply increasing $V_{app}$ back to $-0.5~V$. As we will see in the capacitance data, both reversible and irreversible domain wall motion can lead to NC. However, only reversible domain wall motion leads to stable and hysteresis-free NC behavior, which is needed for device applications. Irreversible domain wall motion leads to hysteresis, energy loss, and reduced device speed, which is detrimental for most applications of NC.

In order to quantify the EOT in the regime of reversible domain wall motion, we compare the $C-V$ relationships obtained from simulations with and without the ferroelectric layer, which are shown in Fig. 3. For a given applied voltage $V_{app}$, the the average charge density $Q$ at the top metal/oxide interface $z = z_{\rm int}$ is calculated as
$Q(z_{\rm int}) = \frac{1}{L_xL_y}\int_0^{L_x}\int_0^{L_y}D_z(x,y,z_{\rm int})dxdy $
where $L_x$ and $L_y$ are lateral dimensions of the film and $D_z(x,y,z_{\rm int}) = \epsilon_0\epsilon_{DE}\times E_{z_{\rm int}} + P_z$ is the normal component of the displacement field in the $x-y$ plane at $z = z_{\rm int}$. The C-V curves are then obtained as $C = dQ/dV_{app}$ from the Q-V relationships. 
Figure~\ref{fig:3} shows that in the lower voltage range, $C_{\rm{MFISM}}$ is much larger than $C_{\rm{MISM}}$, which can only be explained by an effectively NC of the ferroelectric layer, which is electrically in series to the other layers. 
By comparing the MFISM C-V curve with reference MISM curves, we can determine the effective EOT of the gate stack. A $5.5~\rm{\AA}$ EOT is extracted with physical thickness of $8.0~\rm{\AA}$ of the SiO$_2$ interfacial layer in this particular HfO$_2$/ZrO$_2$-based gate stack with a t/o-phase mixture and partially in-plane polarization. Without NC, the EOT would always be larger than the physical SiO$_2$ thickness. However, we can also see from Fig. 3, that for $V_{app} < -0.7~V$ there is a substantial increase in capacitance due to the previously discussed irreversible coalescence of domains shown in Fig. 2(c). Since a small change $\Delta V_{app}$ in this voltage range leads to a large change in charge $\Delta Q$ due to irreversible polarization switching, the calculated capacitance $C = \Delta Q / \Delta V_{app}$ shows a `jump' in Fig. 3. However, upon completion of domain coalescence at $V_{app} < -0.8~V$, there is a marked reduction in overall capacitance due to the cessation of the transient NC effect resulting from the switching process. Therefore, for NC device applications, the operating voltage should be kept low enough that such irreversible switching events do not degrade the overall device performance. For this reason, we limit our discussion on NC from reversible domain wall motion in the following sections i.e. for $V_{app} > -0.7~\rm{V}$.

\subsection{Effect of Polar Axis Angle}
\begin{figure}[h]
    \centering
    \includegraphics[width=\linewidth]{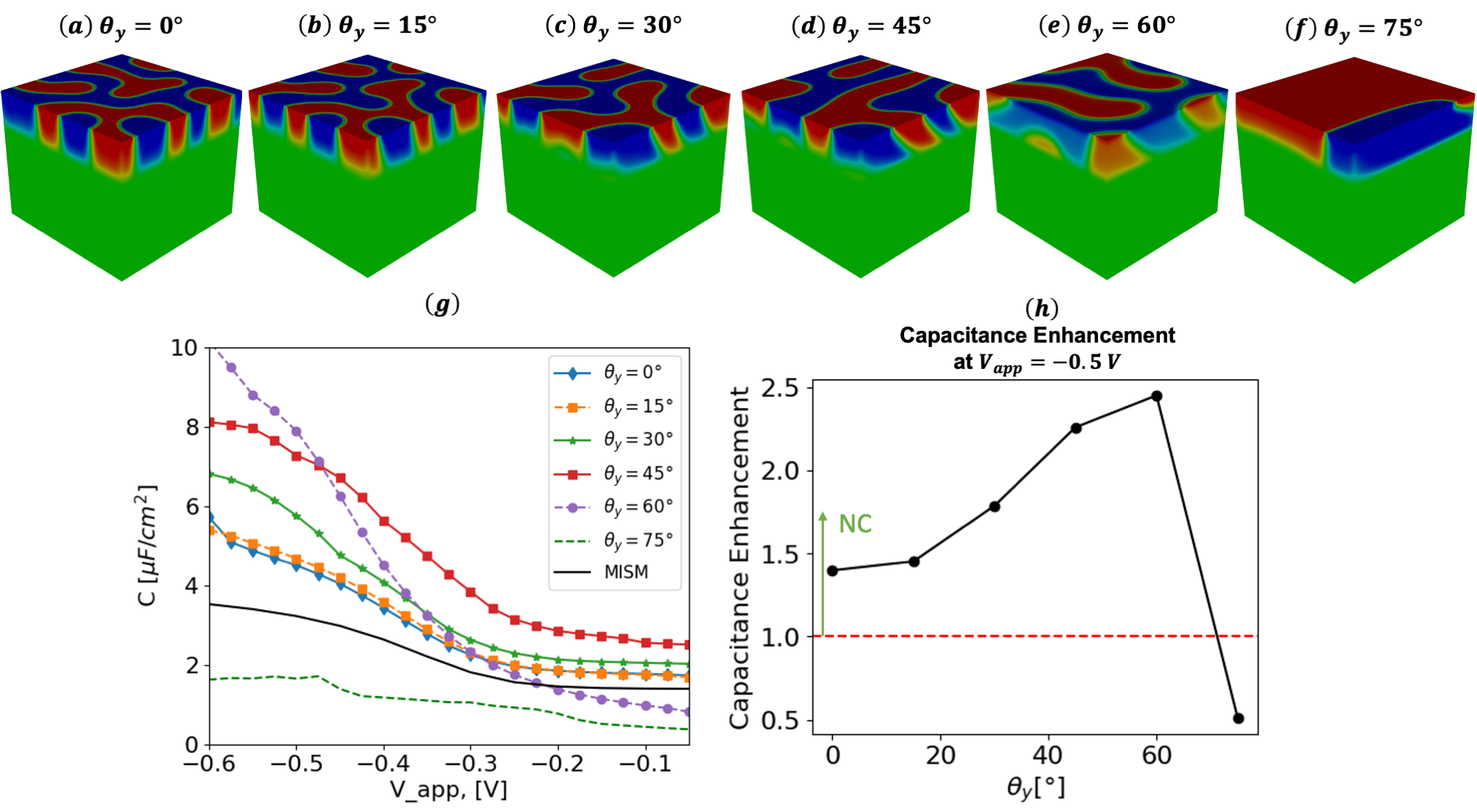}
    \caption{Effect of varying the polar axis angle $\theta_y$. 3D distribution of polarization in the orthrhombic grain shows domain patterns for $V_{app} = 0.0$ V. Making the polarization more in-plane leads to an increase in the domain period until the ferroelectric enters a single-domain state at  $\theta_y \geq 75\degree$. C-V curves for different cases shown in (g) indicate the presence of NC for $\theta_y \leq 60\degree$ with the maximum capacitance enhancement observed at $\theta_y = 60\degree$ for $V_{app} = -0.5$ V.}
    \label{fig:4}
\end{figure}
          
While we previously investigated a fixed polar axis angle of $45\degree$ to reflect a (111) texture of the ferroelectric o-phase, not all grains in these polycrystalline films will have the same polar axis angle. Therefore, in a next step, we focus on the effect of varying the polar axis angle of a single ferroelectric grain on the 3D domain structure and capacitance enhancement due to NC. While keeping all the physical and numerical setup identical as in Figure~\ref{fig:2}, we progressively increase the polar angle $\theta_y$ from $0\degree$ to $75\degree$, similar to what has been previously shown in 2D phase field simulations \cite{Hoffmann2022iedm}. Figure~\ref{fig:4} shows that as $\theta_y$ increases, the polarization is more in-plane which leads to an increase in the domain sizes. A further increase of the polar axis angle to $75\degree$ results in an almost single domain ferroelectric, where domain wall motion
, and thus NC, is not possible. $C-V$ relationships for different $\theta_y$ shown in Figure~\ref{fig:4}(g) show higher capacitance compared to the baseline MISM case for every $\theta_y$ up to $60\degree$, whereas it is lower than the MISM case for $\theta_y = 75\degree$ due to the absence of NC effects. Furthermore, we calculate Capacitance Enhancement as the ratio $C_{\rm{MFISM}} / C_{\rm{MISM}}$ at $V_{app} = -0.5$ V which is a reasonable operating voltage for a NC device. Our results show that the Capacitance Enhancement increases with $\theta_y$ and peaks at $\theta_y = 60\degree$ as shown in Figure~\ref{fig:4}(h). This is attributed to the fact that, as long as the ferroelectric is in a multi-domain state where reversible domain wall motion is possible, a larger domain period increases the stray field energy leading to a larger NC effect \cite{Hoffmann2022iedm}. In addition, the width of the domain walls for the $60\degree$ case is larger compared to the others, which might further contribute to the NC effect by increasing the anisotropy term in the total free energy of the ferroelectric. Therefore, achieving a polar axis angle of the ferroelectric grains close to $60\degree$ would be beneficial to further lower the EOT in future NC devices using ultra-thin HfO$_2$- and ZrO$_2$-based ferroelectrics. 

\subsection{Effect of Ferroelectric Grain Size}
\begin{figure}[h]
    \centering
    \includegraphics[width=0.95\linewidth]{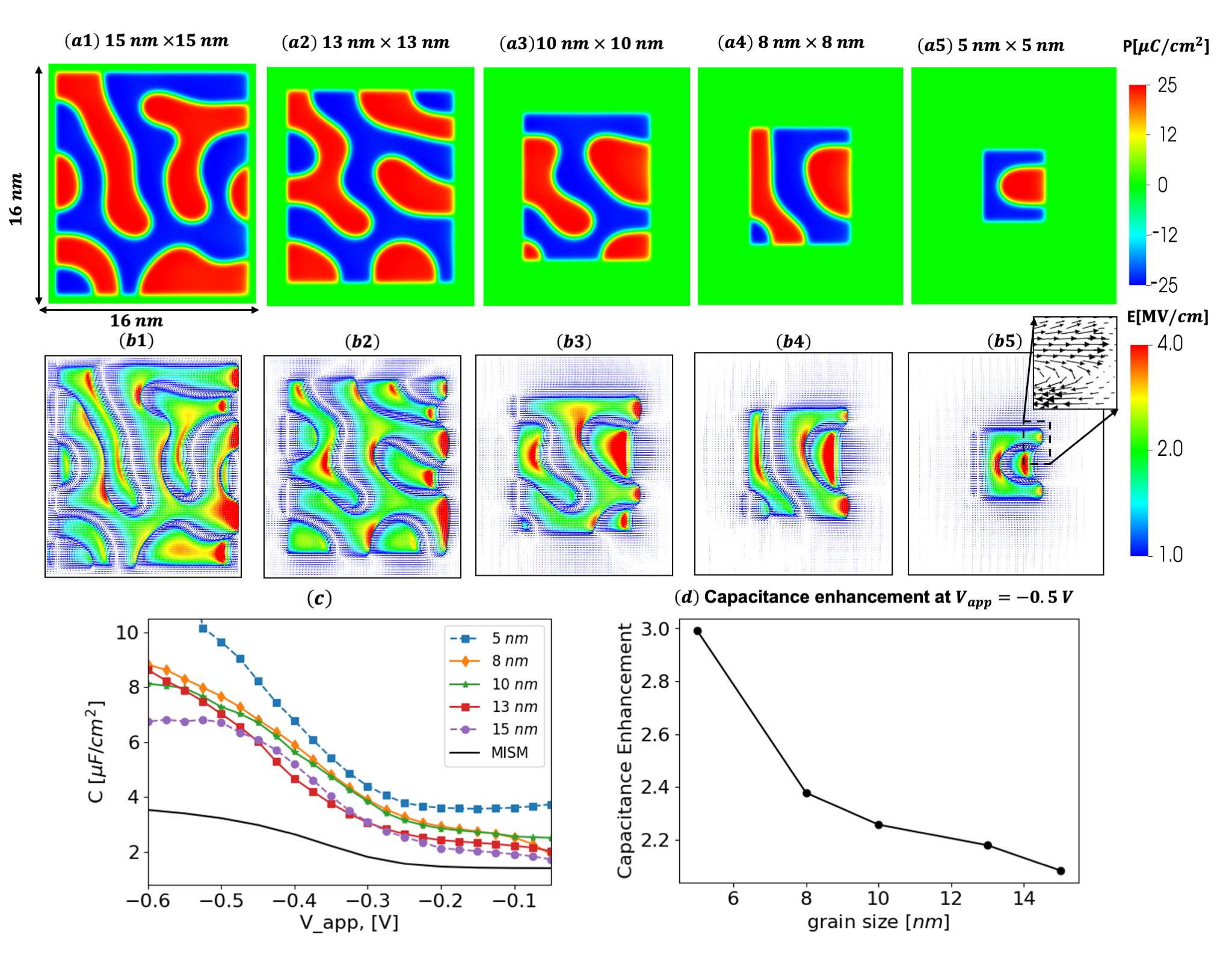}
    \caption{Top view of the polarization domain patterns, electric field, and C-V characteristics for different ferroelectric grain sizes. (a1 - a5) show the domains at the top surface of the ferroelectric layer for grain sizes in 15 nm to 5 nm range, while the total film size remains fixed at 16 nm $\times$ 16 nm. (b1 - b5) show the corresponding vector plots of electric field. (c) shows the C-V curves for these cases together with the baseline MIS case (solid black curve). All the grains sizes simulated show higher capacitance than the baseline case due to NC effects. (d) Capacitance at $V_{app} = -0.5$ V increases as the grain size decreases and is three times the baseline MISM capacitor for a 5 nm $\times$ 5 nm grain.}
    \label{fig:5}
\end{figure}

Another important property of ultra-thin ferroelectric films is the grain size distribution. HfO$_2$- and ZrO$_2$-based polycrystalline ferroelectrics typically have a distribution of lateral grain sizes ranging from about 5 - 50 nm \cite{Lombardo2021,chen2022}. To understand the effect of the grain size on the 3D domain structure and capacitance enhancement we vary the o-phase grain size while keeping all other parameters fixed to the values listed in Table 1. $\theta_y = 45\degree$ is used to capture the preferential (111) texture observed in the experiments, which results in tilted domains in the (x,z) plane (similar to Figure~\ref{fig:2}) for all grains sizes. The top row of Figure~\ref{fig:5} shows the distribution of polarization in the (x,y)-plane at the top surface of the ferroelectric film for grain sizes ranging from 15 nm to 5 nm and an external voltage $V_{app} = 0$ V. Since the lateral size of the film is fixed at 16 nm $\times$ 16 nm, the t-phase fraction increases as the o-phase grains get smaller. However, we only calculate the normalized capacitance for the ferroelectric grain area to exclude the effect from the changing t/o-phase fraction of the film. This means that resulting capacitance enhancements are only representative of o-phase grains embedded in a t-phase matrix. Since interactions between different o-phase grains are not included here, these results might not be valid for films with very high o-phase fractions. 

As can be seen from the polarization distributions in Fig. 5(a1) to (a5), since the domain sizes stay similar, the overall number of domains in one ferroelectric grain is reduced for smaller grain sizes. For a 5 nm $\times$ 5 nm grain, only two domains remain. The electric field distributions in Fig. 5(b1) to (b5) show that the largest fields mostly appear close to the t/o-phase boundaries along the x-axis. Furthermore, the stray fields surrounding the ferroelectric grain extend substantially into the non-polar t-phase regions with a characteristic length on the order of the domain size. When looking at the resulting $C-V$ curves in Fig. 5(c), one can see that all grain sizes show strong capacitance enhancement and thus NC. However, as shown in Fig. 5(d), smaller ferroelectric grains exhibit increasing capacitance enhancement. This increased enhancement for smaller grains might be explained by the reduction in grain volume compared to the volume that is occupied by the stray fields. The energy stored in the stray fields surrounding the ferroelectric grain becomes more and more important for smaller grain sizes. The characteristic length of the stray fields around the ferroelectric grain is getting closer to the grain size itself. At the same time, the number of domains is reduced for smaller grains, which is similar to the trend of increasing polar axis angles until $60\degree$ in Fig. 4(h). This reinforces the notion that fewer domains (as long as there at least two) per grain lead to larger capacitance enhancement and thus lower EOT \cite{Hoffmann2022iedm}. So when building NC transistors, one should try to optimize the ferroelectric film morphology towards smaller grains with polar axis angles close to 45-60$\degree$ to achieve lower EOT.

\subsection{Effect of Gradient Energy Coefficients}

\begin{figure}[h]
    \centering
    \includegraphics[width=\linewidth]{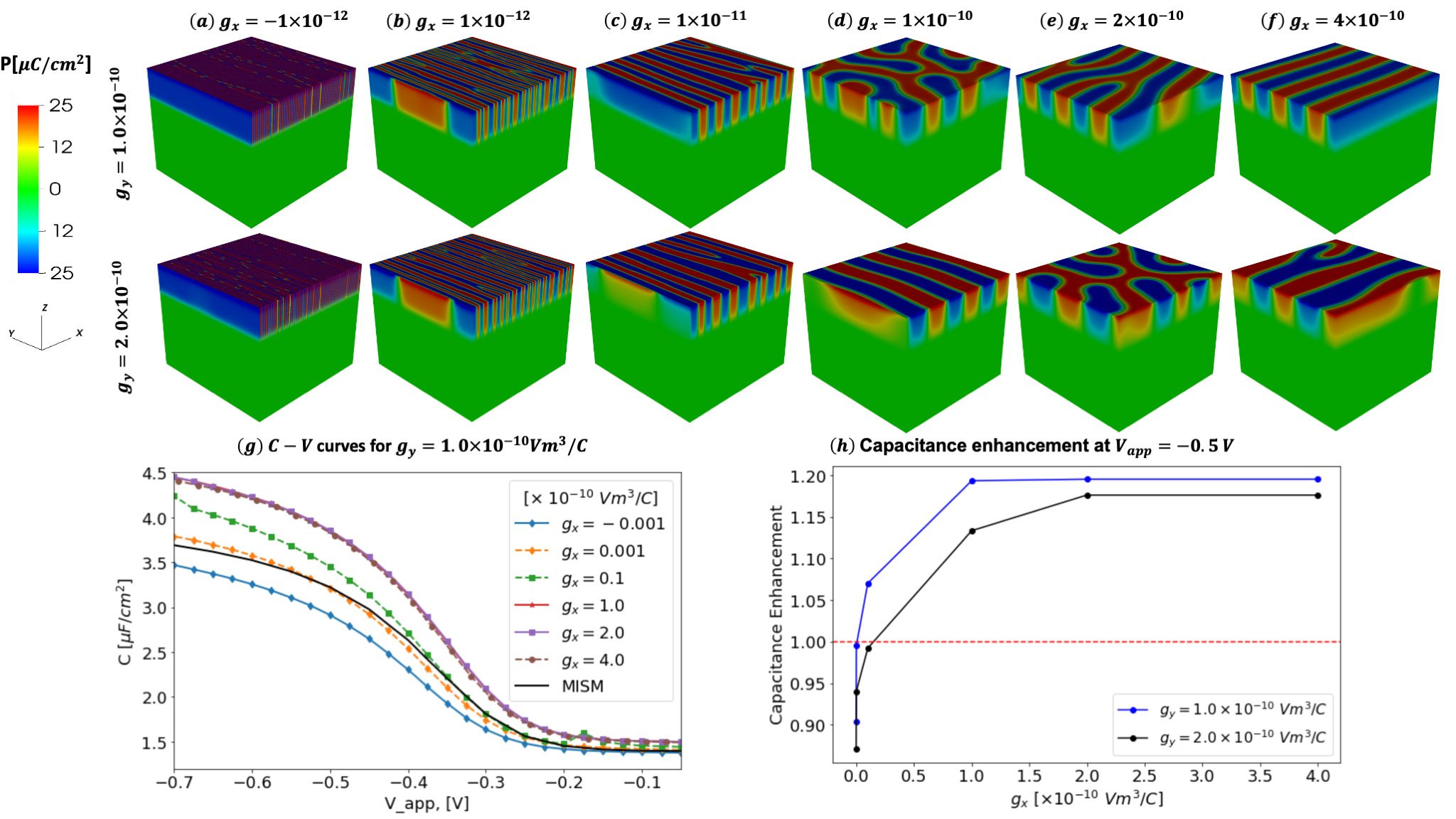}
    \caption{Effect of varying the gradient energy coefficients $g_x$ and $g_y$. (a)-(f) 3D distributions of ferroelectric polarization shows domain patterns for $V_{app} = 0$ V. The top row shows domain patterns for a fixed $g_y = 1\times 10^{-10}$ $~\rm{Vm^3/C}$, while the bottom row uses $g_y = 2\times 10^{-10}$ $~\rm{Vm^3/C}$. (g) Capacitance-voltage (C-V) curves for different $g_x$ and $g_y = 1\times 10^{-10}$ $~\rm{Vm^3/C}$ compared to the reference metal-insulator-semiconductor-metal (MISM) structure. (h) Capacitance enhancement compared to the MISM case at $V_{app} = -0.5$ V for different $g_x$ and $g_y$.}
    \label{fig:6}
\end{figure}

The change in energy related to polarization gradients and domain walls plays a crucial role in the formation of domains, their dynamics, and related NC effects \cite{Saha2019MultiDomainNC,Hoffmann2018}. Although we have so far assumed isotropic gradient coefficients (see Table 1) and thus the same domain coupling in both directions perpendicular to the polar axis, this is not necessarily true for all ferroelectrics. In a recent study \cite{Paul2022DirectionDependentLD}, Paul et al.~used DFT to show that the gradient coefficients in Hf$_{0.5}$Zr$_{0.5}$O$_2$-based ferroelectrics are direction-dependent and exhibit significantly different characteristics along the two axes (x and y) perpendicular to the polar axis (z) of the orthorhombic crystal. Previous 2D phase field simulations cannot fully model this anisotropy in gradient energy coefficients \cite{saha2020multi,Hoffmann2022iedm}. Therefore, we perform 3D phase field simulations to capture the domain structures and dynamics due to strongly anisotropic domain wall coupling in HfO$_2$ and ZrO$_2$ based ferroelectrics and quantify its effect on the domain formation and capacitance enhancement due to NC. For the gradient energy coefficient along the y-direction $(g_y)$ we used a constant value while varying the gradient coefficient $(g_x)$ along the x-direction. To isolate the effect of anisotropic domain wall coupling, we consider a 10 nm $\times$ 10 nm entirely ferroelectric film (i.e. without t-phase) and a fully out-of-plane polar axis. All other parameters are the same as in Table 1. Previous DFT calculations \cite{Lee2020ScalefreeFI} have suggested that HfO$_2$-based materials exhibit ferroelectricity down to the unit cell limit due to slightly negative domain wall coupling, i.e. a preference for the antipolar Pbca phase \cite{cheng2022}. However, 2D phase field simulations have shown such negative domain wall coupling does not lead to NC \cite{Hoffmann2022iedm}, which presents a seeming contradiction when trying to explain the experimentally observed NC effects in HfO$_2$/ZrO$_2$-based materials \cite{cheema2022ultrathin,Jo2023NegativeDC}. To investigate if a negative domain coupling along only one axis and positive coupling along the two other axes can lead to NC in a fully 3D simulation, we varied $g_x$ from a small but negative value ($g_x = -1\times 10^{-12}~\rm{Vm^3/C}$ \cite{Paul2022DirectionDependentLD}) until a relatively large value of $4\times 10^{-10}~\rm{Vm^3/C}$.

Fig. 6(a) to (f) shows the domain patterns for different gradient coefficients $g_x$, while the top and bottom row correspond to two different $g_y$ values. A negative $g_x$ ($g_x = -1.0\times 10^{-12}~\rm{Vm^3/C}$) leads to antipolar ordering resulting in domain periods along x-direction on the scale of ferroelectric unit cell size ($\sim$5 \AA). This unit cell-based local polarization profile results in creation of abrupt and hard domain walls, which cannot reversibly move along the x-axis \cite{Saha2021_JAP}. Furthermore, any change in the overall energy due to domain wall movement in the y-direction is small, because of the small domain period along the x-axis. Therefore, as shown in Fig. 6(g), no capacitance enhancement and thus no NC is observed for a negative $g_x$, even if $g_y$ is positive. As $g_x$ is increased (Fig. 6(b-f)), the domain sizes along the x-direction increase and the domain walls become softer. When $g_x$ is smaller than $g_y$, domains along the y-axis are wider and have softer domain walls resulting in striped domains along the x-axis. When $g_x = g_y$, there is an isotropic domain wall coupling resulting in a more meandering domain pattern instead of stripes (top row Figure~\ref{fig:6}(d) and bottom row of Fig. 6(e)). Further increase in $g_x$ ($g_x > g_y$) results in larger domains along the x-axis, which eventually results in striped domain along the y-axis, see Fig. 6(f). Interestingly, when looking at the C-V curves in Fig. 6(g), the capacitance of the MFISM stack increases with increasing $g_x$ until $g_x = g_y$ and then stays constant. This suggests that the capacitance enhancement due to NC is only limited by the smallest gradient coefficient perpendicular to the polar axis. In Fig. 6(h), it becomes clear that the point where the capacitance enhancement saturates with $g_x$ exactly corresponds to $g_y$. This makes intuitive sense, since the smallest gradient coefficient determines the minimum domain period, which in turn limits the energy that is stored in the stray electric fields that compensate the bound polarization charges at the ferroelectric/dielectric interface. When the domain period along any axis (x or y) is reduced, the NC effect becomes smaller resulting in a lower capacitance enhancement.

The results in Fig. 6 imply, that NC cannot be observed if either $g_x$ or $g_y$ is negative, since the domain period will be on the scale of the unit cell. Therefore, there seems to be a contradiction between the experimental results of NC in HfO$_2$/ZrO$_2$-based ferroelectrics \cite{cheema2022ultrathin,Jo2023NegativeDC}and recent DFT calculations \cite{Lee2020ScalefreeFI,Paul2022DirectionDependentLD}. However, we argue that this discrepancy can be resolved by considering the presence of topological domain walls in HfO$_2$/ZrO$_2$-based materials \cite{Choe2021UnexpectedlyLB}. While the equilibrium domain patterns in these materials might be consistent with a negative gradient energy coefficient $g_x$, their dynamics seem to be affected by topological considerations. Through the application of an electric field, the immobile narrow domain walls caused by a negative gradient energy can be transformed into topological domain walls with high mobility and effective positive coupling \cite{Choe2021UnexpectedlyLB}. This possible transformation of domain wall topology by electric fields seems to lead to an effective positive gradient energy coefficient that facilitates NC effects and can thus explain the observed capacitance enhancement. Therefore, our results indicate that topological domain walls are indeed necessary for explaining experimental NC effects in ultra-thin HfO$_2$/ZrO$_2$-based ferroelectrics.

\section{Conclusion}

We have studied the domain formation and dynamics in ultra-thin mixed-phase ferroelectric films using 3D phase field simulations to better understand experimentally observed NC effects and capacitance enhancement. We have shown that reversible and irreversible domain wall motion can give rise to stable NC and hysteretic switching, respectively, depending on the magnitude of the applied voltage. Through the study of ferroelectric grains embedded in a matrix of a non-ferroelectric phase, we showed that smaller ferroelectric grains and a larger polar axis angles up to 60$\degree$ with respect to the out-of-plane direction leads to stronger capacitance enhancement as long as at least one domain wall inside the grain can reversibly move. Furthermore, we studied the effect of anisotropic domain wall coupling in ultra-thin HfO$_2$/ZrO$_2$-based ferroelectrics and found that a negative gradient energy even along one axis impedes NC. Only the smallest gradient energy coefficient determines the capacitance enhancement. This suggests that topological domain walls in HfO$_2$/ZrO$_2$-based ferroelectrics lead to effecitvely positive domain wall coupling, which is necessary to explain experimentally observed NC effects.


\section{Computational Section}
All the simulations reported in this work have been carried out using the 3D phase-field framework, FerroX~\cite{Kumar2023}. The distribution of electric potential in the system is obtained by solving Poisson's equation in the following form:
\begin{equation}
    \nabla\cdot\epsilon\nabla\Phi = \nabla\cdot \mathbf{P} - \rho
\end{equation}
where $\epsilon$ is a spatially-varying permittivity and $\rho$ is the total free charge density in the semiconductor region. $\rho$ depends on the local distribution of electric potential and is computed using the following equation:
\begin{equation}
    \rho(\mathbf{r}) = e\left[n_p - n_e + N_d^+ - N_a^-\right]
\label{eq:rho}
\end{equation}
where $e$ is the elementary charge and $n_p(\mathbf{r})$, $n_e(\mathbf{r})$, $N_d^+(\mathbf{r})$, and $N_a^-(\mathbf{r})$ are densities of holes, electrons, ionized donors, and acceptors at spatial location $\mathbf{r}$. We use a $p-$doped silicon as the semiconductor with an acceptor doping of $10^{15}~cm^{-3}$ where the distribution of electrons and holes is estimated as a function of $\Phi$ using Maxwell-Boltzmann statistics. The evolution of polarization in the ferroelectric polycrystal is described by the time-dependent Ginzburg–Landau (TDGL) equation:
\begin{equation}
    \frac{\partial\mathbf{P}(\mathbf{r},t)}{\partial{t}} = -\Gamma\frac{\delta F}{\delta\mathbf{P}(\mathbf{r},t)}
\label{eq:TDGL1}
\end{equation}
where $\mathbf{P}$ is the electric polarization vector, $\mathbf{r} = (x,y,z)$ is the spatial vector, $\Gamma$ is the kinetic or viscosity coefficient, and $F$ is the total free energy of system as a function of $\mathbf{P}(\mathbf{r},t)$ and takes into account the contributions due to the bulk Landau free energy, the domain wall energy or gradient energy, and the electric energy of the applied electric field. 
Exact forms of these energy densities are elaborated in ~\cite{Kumar2023}. 

In order to model the polycrstalline ferroelectric we introduce a common global coordinnate system $(x,y,z)$ for all grains. In the t-phase, polarization is consistently zero. However, within each o-phase grain, it exhibits inhomogeneous variations, represented using local coordinates $(p,q,r)$ specific to each grain. These local coordinates are determined by polar axis angles $(\theta_x, \theta_y, \theta_z)$, which correspond to three counter-clockwise rotations with respect to the global coordinates $(x, y, z)$. The rotaion matrix from the global to local coordinate system is given by:

\begin{equation*}
    R = \begin{bmatrix}
            cos\theta_ycos\theta_z & sin\theta_xsin\theta_ycos\theta_z - cos\theta_xsin\theta_z & cos\theta_xsin\theta_ycos\theta_z + sin\theta_xsin\theta_z\\
            cos\theta_ysin\theta_z & sin\theta_xsin\theta_ysin\theta_z + cos\theta_xcos\theta_z & cos\theta_xsin\theta_ysin\theta_z - sin\theta_xcos\theta_z \\
            -sin\theta_y & sin\theta_xcos\theta_y & cos\theta_xcos\theta_y
        \end{bmatrix}
\end{equation*} 
 Accordingly, the first and second order derivative of a scalar valued function $f$ in the local coordinates are obtained from the corresponding terms in the global coordinates by 
 \begin{equation*}
     \begin{bmatrix}
         f_p \\ f_q \\ f_r
     \end{bmatrix}
     = R
     \begin{bmatrix}
         f_x \\ f_y \\ f_z
     \end{bmatrix}
 \end{equation*} 

\begin{align*}
    f_{pp} &= R_{11}^2f_{xx} + R_{12}^2f_{yy} + R_{13}^2f_{zz}
             +2(R_{11}R_{12}f_{xy} + R_{12}R_{13}f_{yz} + R_{13}R_{11}f_{xz})\\
    f_{qq} &= R_{21}^2f_{xx} + R_{22}^2f_{yy} + R_{23}^2f_{zz}
             +2(R_{21}R_{22}f_{xy} + R_{22}R_{23}f_{yz} + R_{23}R_{21}f_{xz})\\
    f_{rr} &= R_{31}^2f_{xx} + R_{32}^2f_{yy} + R_{33}^2f_{zz}
             +2(R_{31}R_{32}f_{xy} + R_{32}R_{33}f_{yz} + R_{33}R_{31}f_{xz})
\end{align*}

We discretize this coupled system of partial differential equations using a finite difference approach, with an overall scheme that is second-order accurate in both space and time. In order to achieve a massively parallel manycore/GPU implementation of our structured grid simulations, we leverage the DOE Exascale Computing Project (ECP) code framework, AMReX, developed by Zhang et al.~\cite{zhang2021amrex}. Additional details on numerical approach, parallelization, and scaling studies are given in \cite{Kumar2023}.

\medskip
Each simulation begins with $P_z$ initialized from a uniform distribution of random numbers in the interval $[-0.2, 0.2]~\mu C/cm^2$ and a corresponding self-consistent $\Phi(\vec{r})$ and $\rho(\vec{r})$. 
The voltage applied across the stack is controlled by specifying the electric potentials at the metal layers as the boundary condition for Poisson's equation. We start with an initial voltage of $0~V$ and gradually decrease it by $0.05~V$ after achieving a steady state, repeating this process until we establish the desired $Q-V$ relationship within the specified voltage range.


\medskip
\textbf{Data availability} \par
The datasets generated during and/or analyzed during the current study are available in the zenodo repository at~\cite{kumar_2024_10632037}.

\medskip
\textbf{Acknowledgements} \par 
This work was supported by the US Department of Energy, Office of Science, Office of Basic Energy Sciences, the Microelectronics Co-Design Research Program, under contract no. DE-AC02- 05-CH11231 (Codesign of Ultra-Low-Voltage Beyond CMOS Micro- electronics) for the development of design tools for low-power microelectronics. This research used resources of the National Energy Research Scientific Computing Center, a DOE Office of Science User Facility supported by the Office of Science of the U.S. Department of Energy under Contract No. DE-AC02-05CH11231. This research leveraged the open source AMReX code, https://github.com/AMReX-Codes/amrex. We acknowledge all AMReX contributors.

\medskip


\bibliographystyle{MSP}
\bibliography{main}

\end{document}